\documentclass[aps,twocolumn]{revtex4}

\begin{document}

\def\bb{\begin{equation}}
\def\ee{\end{equation}}

\title{Hidden glassy behaviors in an ideal Heisenberg Kagom\'{e}
antiferromagnet}
\author{J.D. Lee}
\address{W.M. Keck Laboratories, California Institute of Technology,
Pasadena, CA 91125}
\date{\today}

\begin{abstract}
Dynamics of classical Heisenberg spins, ${\bf S}_i=({\bf s}_i,S_{iz})$,
on the Kagom\'{e} lattice has been studied.
An ideal Heisenberg Kagom\'{e} antiferromagnet is 
known to remain disordered down to $T=0$
due to the macroscopic degeneracy of the ground state. Through the study, 
however, we find that ${\bf S}_i$ and their planar components ${\bf s}_i$
behave in a qualitatively different way and especially 
planar spins (${\bf s}_i$) show the exotic glass-like transition in the very
low temperature $T\sim 0.003J$ ($J$: spin exchange).
The glassy behaviors of ${\bf s}_i$ would be found to be 
driven by the spin-nematic fluctuations, different from ordinary spin glasses
by disorders or anisotropies.
\end{abstract}

\pacs{75.40.Gb, 75.10.Nr, 75.10.Hk, 75.50.Ee}

\maketitle
Magnetic systems in which the exchange energy is not minimized 
by a simple regular arrangement of the spins are called
the frustrated magnetic systems.
Frustrated antiferromegnets induced by the geometry of the lattice
have been a recent subject attracting much experimental and theoretical 
interests\cite{Schiffer}. One example of such systems is the antiferromagnet
on the Kagom\'{e} lattice. It consists of triangles.
The Kagom\'{e} lattice is, however, more frustrated than 
the triangular lattice because triangles on the former lattice
share only one vertex, but on the latter share one side.
Indeed, classical Heisenberg spins on the Kagom\'{e} lattice
have an infinitely degenerate ground state and some novel ground
state is expected as $T\to 0$ without developing long-range order
at any $T$, which has been a main reason drawing considerable
attention\cite{Huse,Harris,Chalker,Ritchey,Reimers}.
Their experimental realizations are, for instance, 
the insulating layered compound SrCr$_{8-x}$Ga$_{4+x}$O$_{19}$ 
(Cr$^{3+}$ carries $S=3/2$)\cite{Ramirez,Broholm,Uemura}
and the jarosites $A$Fe$_3$(OH)$_6$(SO$_4$)$_2$ ($A$=K, H$_3$O, Na, NH$_4$,
and so forth; Fe$^{3+}$ carries $S=5/2$)\cite{Inami,Wills}.

\begin{figure}
\vspace*{9.5cm}
\includegraphics{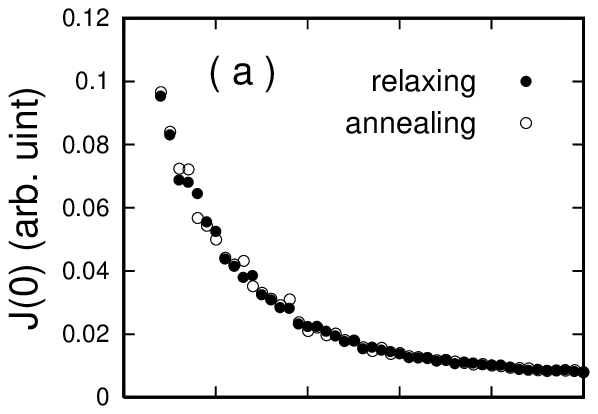}
\caption{Local spin-spin correlation functions at $\omega=0$
with respect to temperatures $T$ for both cases of relaxing and annealing.
(a) gives behaviors of $J(0)$ which is history-independent,
whereas (b) shows that $j(0)$ is history-dependent
approximately below $T\sim 0.003J$. The inset in (a) is the spin
configuration for the $\sqrt{3}\times\sqrt{3}$ state and the inset in (b)
is the profile of $J(0)-j(0)$, i.e. $J_z(0)$ with $T$ (the line is a guide
for the eye).
}
\label{FIG1}
\end{figure}

Extremely high degeneracy of the ground state inhibits
the long-range order at any $T$. But it has been known that thermal 
fluctuations in the system select a subset of ground state
manifold at low $T$ (known as ordering by disorder)\cite{Chalker,Reimers}, 
even if this subset is not clear enough to lead to long-range 
N\'{e}el order. Chalker {\it et al.}\cite{Chalker} 
have argued by low temperature
expansions that thermal fluctuations resolve degeneracy of ground states
and cause the system to select a coplanar nematic ground states.
There are at least two coplanar ground states
possessing long-range N\'{e}el order; one is the so-called 
$q=0$ state and the other is $\sqrt{3}\times\sqrt{3}$ state (See the
inset in Fig.\ref{FIG1}(a); in the $q=0$ state, spins along lines
are placed in an alternating sequence compared to 
the $\sqrt{3}\times\sqrt{3}$ state). Reimers and Berlinsky\cite{Reimers}
have shown, through the Monte Carlo simulation, that an increase in
the order parameter of the $\sqrt{3}\times\sqrt{3}$ state is dominant
over the $q=0$ state. In addition, Sachdev\cite{Sachdev}
has also argued that the $\sqrt{3}\times\sqrt{3}$ state is
further accepted and favored in the low temperature limit
by quantum fluctuations.

Magnetic frustration used to be found with the site disorder,
even if these two features may appear independently. When both,
disorder and frustration, are strong, its interplay gives 
place to the spin glass phase\cite{Chowdhury}. On the other hand,
the antiferromagnetic Heisenberg model on the Kagom\'{e} lattice
is an ideal geometrically frustrated magnet without disorders.
Nevertheless, in some of prototype systems of the Kagom\'{e} lattice,
the spin glass ordering at a finite temperature is actually observed
because of deviations from an ideal one.
In SrCr$_{8-x}$Ga$_{4+x}$O$_{19}$, $T_g=3.3$ K is found in a case of $x=0$ 
corresponding to 89\% occupation of Cr 
sites\cite{Ramirez,Broholm,Uemura}. In the system, Cr$^{3+}$
just enters three different octahedral sites (namely, 2a, 12k, 4f)
with an almost random distribution and SrCr$_8$Ga$_4$O$_{19}$
has then $\sim$14\% site disorder (nonmagnetic impurities) 
on the Kagom\'{e} layer (12k sites).
The glass order is enhanced by the additional dilution of 
Cr sites\cite{Ramirez}. The glassy phase in nondisordered 
frustrated magnetic systems has been attracting much interests.
Most two-dimensional magnetic materials exhibit some kinds of
spin anisotropy. A critical slowing down of magnetic fluctuations is
found in the hydronium-jarosite (H$_3$O)Fe$_3$(OH)$_6$(SO$_4$)$_2$
with much less disorders\cite{Wills}. The glass phase in the system
may be a bit different from in SrCr$_8$Ga$_4$O$_{19}$,
due to the $xy$ anisotropy ("easy-plane"), which could drive the Kagom\'{e}
antiferromagnet to experience the topological 
Kosterlitz-Thouless-type transition to a glassy state\cite{Ritchey}.
The glassy behavior in the Kagom\'{e} antiferromagnet with
the easy-axis type anisotropy has been also explored\cite{Kuroda}. 

In this paper, through the study of dynamical properties, 
we report the hidden glassy behaviors of Heisenberg spins 
on the ideal Kagom\'{e} lattice without any disorder or anisotropy.
Keren\cite{Keren} has investigated the stability of the spin order 
against excitation and the dynamical responses of spins on
the Kagom\'{e} lattice. Actually, in his study, there is found no 
transition down to $T=0$ as there should not be.
Nevertheless, we find that behaviors of ${\bf S}_i$ and its planar component 
${\bf s}_i$ are qualitatively different from each other
and ${\bf s}_i$ undergoes a glassy transition around $T\sim 0.003J$
unlike ${\bf S}_i$.

Taking the classical Heisenberg spins as ${\bf S}_i=({\bf s}_i,S_{iz})$, 
we have the Hamiltonian
\bb\label{eq1}
{\cal H}=J\sum_{ij}{\bf s}_i\cdot{\bf s}_j+J\sum_{ij}S_{iz}S_{jz},
\ee
where ${\bf s}_i^2+S_{iz}^2=1$. 
Reimers and Berlinsky\cite{Reimers} have demonstrated that
the tendency toward spin-nematic order starts at $T\sim 0.01J$, which
motivates us to explore the respective dynamics of ${\bf S}_i$ 
and ${\bf s}_i$ at the relevant temperature range, i.e. 
$T\lesssim 0.01J$. Particularly, we have interests in the local 
spin-spin correlation functions as
\begin{eqnarray}\label{eq2}
J(\omega)&=&\frac{1}{N}\int d\tau e^{-i\omega\tau}\sum_i
\langle {\bf S}_i\cdot{\bf S}_i(\tau)\rangle,
\nonumber \\
j(\omega)&=&\frac{1}{N}\int d\tau e^{-i\omega\tau}\sum_i
\langle {\bf s}_i\cdot{\bf s}_i(\tau)\rangle.
\end{eqnarray}
The correlation functions such as $J(\omega)$ and $j(\omega)$ can be
most properly evaluated by the Monte Carlo simulation\cite{Landau}.
The time evolution of spins are
governed by the equation of motion $\partial {\bf S}_i/\partial {\tau}
=-J{\bf S}_i\times\sum_{j\neq i}{\bf S}_j$. A practical integration of
the equation is done using the Suzuki-Trotter decomposition
on the three sublattices (say $A$, $B$, and $C$) of the Kagom\'{e} lattice,
$\{{\bf S}_i(\tau+\delta\tau)\}=e^{(A+B+C)\delta\tau}\{{\bf S}_i(\tau)\}$, where
$e^{(A+B+C)\delta\tau}$ is decomposed using $e^{(X+Y)\delta\tau}=\prod_{i=1}^5
e^{p_iX\delta\tau/2}e^{p_iY\delta\tau}e^{p_iX\delta\tau/2}
+{\cal O}(\delta\tau^5)$ and 
$p_1=p_2=p_4=p_5=p=1/(4-4^{1/3})$ and $p_3=1-4p$\cite{Landau}.
We use the hybrid Monte Carlo procedure which combines 
the Metropolis update and the overrelaxation update\cite{Brown}. 
One hybrid Monte Carlo step in our calculation consists of 
two Metropolis steps and four overrelaxation steps. 
Using the hybrid algorithm, ${\cal O}(10^4)$ hybrid steps are typically used 
to reach the equilibrium configuration\cite{hybrid}.
A system size is defined by the linear dimension $L$ of the lattice
and the total number of spins $N$ is $3L^2$.
In an actual simulation for dynamical correlation functions,
the time integration has been performed up to $\tau_{max}=26214.4J^{-1}$
with a step of $\delta\tau=0.2J^{-1}$. The thermal average of time-dependent 
variables is computed over the observables obtained by evolving all independent
initial equilibrium configurations. Typically, $10^3$ configurations were 
generated at a given $T$. An adopted system size for dynamical calculations
is $L=24$, that is, 1728 spins\cite{Lsize}.

\begin{figure}
\vspace*{5.5cm}
\includegraphics{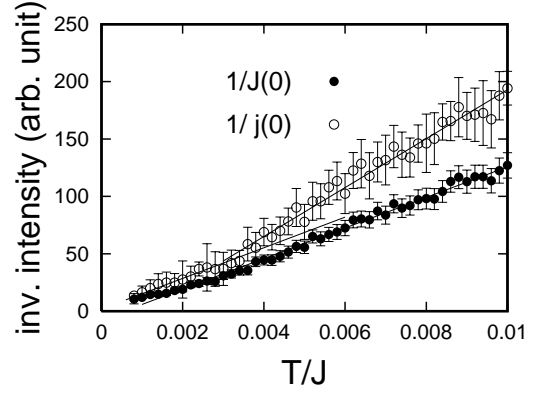}
\caption{Comparison of inverse intensities $1/J(0)$ and $1/j(0)$ 
with temperatures $T$. Inverse intensities are well fitted with 
a linear $T$. $1/j(0)$ shows a clear refraction near $\sim 0.003J$,
which is not in $1/J(0)$.}
\label{FIG2}
\end{figure}

In Fig.\ref{FIG1}, the temperature-dependences of $J(0)$ and $j(0)$ are provided
both when relaxing and annealing the system. $J(0)$ (or $j(0)$)
is a quasi-elastic response of local spin-spin correlation function
and a quantity directly related to the spin-lattice relaxation
rate $T_1$ of a local probe like the nuclear magnetic resonance (NMR). 
For a case of relaxing, the nominal state at $T=0$ is taken as 
the $\sqrt{3}\times\sqrt{3}$ state. For a case of annealing, 
on the other hand, after heating the $\sqrt{3}\times\sqrt{3}$ 
state to $T=0.05J$, we obtain the equilibrium state 
by gradual cooling. Fig.\ref{FIG1} shows that
$J(0)$ is history-independent, whereas $j(0)$ is history-dependent
roughly below $T\sim 0.003J$. History-dependent magnetic responses 
can be one of indications of the glassy phase because of 
the slow dynamics of spins in the phase. 
Hereafter the aimed configuration is obtained by warming the system
from the $\sqrt{3}\times\sqrt{3}$ state, if not stated otherwise.
In the inset of the figure, $J(0)-j(0)$, 
i.e. $J_z(0)$ are given. A much sharper comparison between
$J(0)$ and $j(0)$ is done by taking their inverses. 
In Fig.\ref{FIG2}, $1/J(0)$ and $1/j(0)$ are shown more or less linear 
with respect to $T$ and $1/j(0)$ suffers an abrupt and appreciable change
of the proportionality constant near $T\sim 0.003J$, which 
is not found in $1/J(0)$. 

\begin{figure}
\vspace*{9.5cm}
\includegraphics{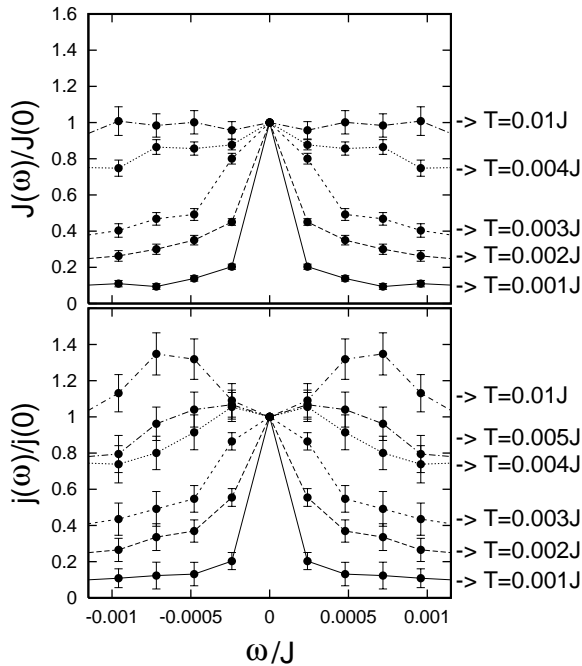}
\caption{$J(\omega)/J(0)$ and $j(\omega)/j(0)$ for small $\omega$'s
($|\omega|\ll 0.003J$) at several temperatures. In $j(\omega)/j(0)$,
qualitative changes of $\omega$-behaviors are observed around
$T\sim 0.003J$.
}
\label{FIG3}
\end{figure}

It should be also interesting to investigate the behaviors of $J(\omega)$ and
$j(\omega)$ in the low energy region ($\omega \ll 0.003J$).
Fischer and Kinzel\cite{Fischer} have extended the dynamical analysis of
early spin glass theories and predicted 
a crossover in low energy behaviors of the ac magnetic susceptibility 
($\chi^{\prime\prime}(\omega)$; basically, same as $J(\omega)$) from 
$\chi^{\prime\prime}(\omega)\sim \omega$ in high temperatures 
to $\chi^{\prime\prime}(\omega)\sim \omega^{\nu(T)}$ near $T_g$ and 
especially $\nu(T_g)=1/2$. Paulsen {\it et al.}\cite{Paulsen} have found that
$\chi^{\prime\prime}(\omega)$ in the spin glass system Eu$_{0.4}$Sr$_{0.6}$S
has a power-law dependence on frequency like
$\chi^{\prime\prime}(\omega)\sim \omega^{\nu(T)}$
and $\nu(T)$ has abruptly changed from $\nu(T\lesssim T_g)\sim 0.09$ to
$\nu(T\gtrsim T_g)\sim 0.4$ around the freezing temperature $T_g$.
Broholm {\it et al.}\cite{Broholm} have also reported the similar power law 
behavior like $\omega^{\nu(T)}$ of the local response function 
in SrCr$_{8-x}$Ga$_{4+x}$O$_{19}$ and changes of $\nu(T)$ around 
$T_g$, that is, $\nu(T)\sim 0$ below $T_g$ and 
$\sim 1$ in high temperatures.
In Fig.\ref{FIG3}, the low energy responses of the spin-spin
correlation functions are given. Behaviors of $j(\omega)$ in
the lower panel of Fig.\ref{FIG3} is appealing in that 
$j(\omega)\sim \omega^0$ ($j(\omega)$ is expected to follow 
$\sim 1/(\omega^2+\Gamma(T)^2)$) below $T\sim 0.003J$ changes to 
$j(\omega)\sim \omega^{\nu}$ ($\nu>0$) above $T\sim 0.003J$.
No similar changes are found in $J(\omega)$ as shown in the upper panel 
of the figure. All previous results of Figs.\ref{FIG1}-\ref{FIG3} 
on $j(\omega)$ are subordinate to the glass-like phase transition 
at $T\sim 0.003J$ and quite consistent with each other. On the contrary,
any indication of spin phase transition down to $T\to 0$ is not found
in magnetic responses of ${\bf S}_i$ (i.e. $J(0)$ and $J(\omega)$),
which is consistent with Keren's\cite{Keren}.

\begin{figure}
\vspace*{4.cm}
\includegraphics{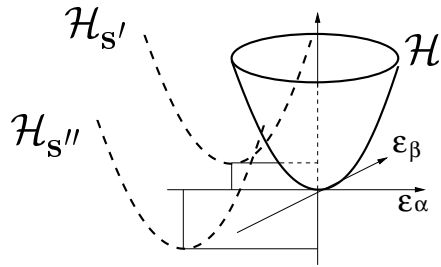}
\caption{Figurative diagram of the Hamiltonian (${\cal H}$)
in the low-temperature expansion and its planar counterpart 
(${\cal H}_{{\bf s}^{\prime}}$ and ${\cal H}_{{\bf s}^{\prime\prime}}$).
${\cal H}_{{\bf s}^{\prime}}$ and ${\cal H}_{{\bf s}^{\prime\prime}}$,
projected to the plane of $\epsilon_{\beta}=0$,
are for two different local spin distortions, while ${\cal H}$ has
a stable saddle point for (small) spin distortions.
}
\label{FIG4}
\end{figure}

We now importantly note that ${\bf S}_i$ and ${\bf s}_i$ behave 
in a qualitatively different way. In order to understand the differences,
we pay attention to the low energy sector of the Hamiltonian.
Following Chalker {\it et al.}\cite{Chalker},
${\cal H}$, in the low temperature expansion,
on the Kagom\'{e} lattice is schematically
$$
{\cal H}={\cal H}_0+{\cal H}_{\rm h}(\epsilon_{\alpha}^2,\epsilon_{\beta}^2)
        +{\cal H}_{\rm anh}(\epsilon_{\alpha}\epsilon_{\beta}^2,
          \epsilon_{\beta}^4,...),
$$
where $\epsilon_{\alpha}$ ($\epsilon_{\beta}$) is the fluctuation
perpendicular to ${\bf S}_i$ in the particular planar ground state 
and in (out of) the spin plane and ${\cal H}_{\rm h}$
and ${\cal H}_{\rm anh}$ includes harmonic and anharmonic fluctuations, 
respectively. In harmonic fluctuations (within the linear spin waves), 
$\epsilon_{\alpha}^2$ gives
three transverse modes and $\epsilon_{\beta}^2$ two transverse modes and
one zero mode per unit cell comprising three spins.
The role of quartic potential for the zero mode is then unique and is 
to stabilize the planar states by inducing the spin-nematic fluctuation.
Now, let us think of the planar counterpart of ${\cal H}$, i.e. 
${\cal H}_{\bf s}$,
$$
{\cal H}_{\bf s}={\cal H}_0+{\cal H}_{\rm h}(\epsilon_{\alpha}^2)
        +{\cal H}_{\rm anh}(\epsilon_{\alpha}\epsilon_{\beta}^2,
          \epsilon_{\beta}^4,...).
$$
In ${\cal H}_{\bf s}$, terms of $\epsilon_{\alpha}\epsilon_{\beta}^2$
and $\epsilon_{\beta}^4$ disturb the system 
like disorders (but not static) in a complicated way.
For a weakly (or non-) disordered case, it is known that
the ground state configuration is such that the energy of each
separate triangle is minimized ({\it "rule of satisfied triangles"}),
implying the glass phase will not occur\cite{Shender}.
In ${\cal H}_{\bf s}$, however, in Fig.\ref{FIG4}, the energy minimization
of each triangle depends on the local spin distortions produced by
neighboring triangles, in other words, by correlations between 
fluctuating degrees of freedom (e.g. $\epsilon_{\alpha}^i
{\epsilon_{\beta}^j}^2$ or ${\epsilon_{\beta}^i}^2{\epsilon_{\beta}^j}^2$).
Alternatively, terms of $\epsilon_{\alpha}\epsilon_{\beta}^2$
and $\epsilon_{\beta}^4$ can be also understood to lead to
planar spins of dynamically varied magnitudes, which act as strong
disorders such that glassy behavior may occur.
To integrate out $\{\epsilon_{\alpha}\}$ in the partition function 
or correlation functions leaves terms of $\epsilon_{\beta}^4$.
If we note its role in ${\cal H}$ and the ground state, the glass-like order
of planar spins can be attributed to the spin-nematic fluctuations.
The glass-like order in the system is different from conventional 
spin glasses in that it is originated by the dynamical fluctuations of
inherent degrees of freedom, not by the static physical disorders
or anisotropies in the systems. 

\begin{figure}
\vspace*{5.5cm}
\includegraphics{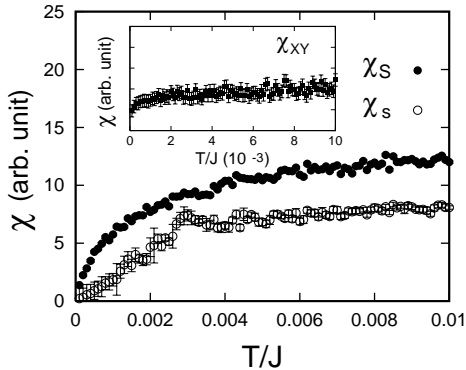}
\caption{Static spin susceptibilities for ${\bf S}_i$ and ${\bf s}_i$,
$\chi_{\bf S}$ and $\chi_{\bf s}$. The inset gives the spin susceptibility
for XY spins, $\chi_{\rm XY}$. Calculations are done for 2700 spins
on the Kagom\'{e} lattice ($L=30$).
}
\label{FIG5}
\end{figure}

Finally, it may be interesting to test
a case of XY spins (planar spins {\it by definition}) 
without such fluctuations\cite{Hxy}, 
for which the easiest way is probably to examine
the static spin susceptibility. It is also a good indicator
of the spin ordering and shows a cusp at the transition temperature
in the spin glass\cite{Chowdhury,Chalupa}. 
The spin susceptibilities are defined for ${\bf S}_i$ and ${\bf s}_i$ as
$\chi_{\bf S}(T)\propto 1/T\sum_{i\neq j}\left[\langle
{\bf S}_i\cdot{\bf S}_j\rangle-\langle{\bf S}_i\rangle\cdot
\langle{\bf S}_j\rangle\right]$ and
$\chi_{\bf s}(T)\propto 1/T\sum_{i\neq j}\left[\langle
{\bf s}_i\cdot{\bf s}_j\rangle-\langle{\bf s}_i\rangle\cdot
\langle{\bf s}_j\rangle\right]$, respectively.
In Fig.\ref{FIG5}, it is found that only $\chi_{\rm s}(T)$
shows a cusp at $T\sim 0.003J$ and other $\chi_{\bf S}(T)$ and 
$\chi_{\rm XY}(T)$ increases smoothly without any singularities 
with respect to $T$, even if all the three cases are started from
the seemingly same spin configuration (i.e. $\sqrt{3}\times\sqrt{3}$ state).
Spin susceptibilities in Fig.\ref{FIG5} ascertain 
the glassy behavior of the planar components of 
Heisenberg spins on the Kagom\'{e} lattice and further support that
it would be induced by the spin-nematic fluctuation.
As mentioned, one of the most outstanding features
of the frustrated magnet is the high degeneracy of the ground state,
whose peculiarity could result in the spin-nematic ordering in the 
limit of very low temperatures ($T\ll J$).
The nematic glassy state should then be crucial and eminently
important in understanding the novel ground state of the frustrated magnet.

In summary, we have found and discussed qualitative differences 
between ${\bf S}_i$ and its planar component ${\bf s}_i$ in an
ideal Heisenberg Kagom\'{e} antiferromagnet. It is remarkable to find
the hidden glass-like order of ${\bf s}_i$ even in the Heisenberg model 
on the Kagom\'{e} lattice without any disorder or anisotropy. 
The glassy behavior has been shown through dynamical spin responses
and static susceptibilities. We find that it would be driven by
the dynamical fluctuation of spin-nematic order, different from
ordinary spin glasses by static disorders or anisotropies.

Discussions with Brent Fultz and Tim Kelley are appreciated.
Technical helps on the parallel computation from Jiao Lin
are appreciated, too. This work was supported 
by the U.S. Department of Energy under contract DE-FG03-01ER45950.


\begin{references}
\bibitem{hybrid} ${\cal O}(10^4)$ hybrid steps look enough to reach the
                 equilibrium state, which usually requires more than 
                 ${\cal O}(10^6)$ nonhybrid Metropolis steps.
\bibitem{Lsize} The lattice size dependence is not so strong 
                that $L=24$ (1728 spins) is found fine (See Ref.\cite{Keren}).
\end{references}
\end{document}